# Discontinuous metric programming in liquid crystalline elastomers

*Tayler S. Hebner,[a] Riley G.A. Bowman,[a] Daniel Duffy,[b] Cyrus Mostajeran,[b,c] Itay Griniasty,[d] Itai Cohen,[d] Mark Warner,[e] Christopher N. Bowman,[a,f] and Timothy J. White\*[a,f]*

[a] Department of Chemical and Biological Engineering, University of Colorado Boulder, 596 UCB, Boulder, CO 80309, USA

[b] Department of Engineering, University of Cambridge, Cambridge, England CB2 1PZ, United Kingdom

[c] School of Physical and Mathematical Sciences, Nanyang Technological University, Singapore 637371

[d] Laboratory of Atomic and Solid State Physics, Cornell University, Ithaca, New York 14853-2501, USA

[e] Cavendish Laboratory, University of Cambridge, Cambridge CB3 0HE, United Kingdom

[f] Materials Science and Engineering Program, University of Colorado Boulder, 596 UCB, Boulder, CO 80309, USA

ABSTRACT Liquid crystalline elastomers (LCEs) are shape-changing materials that exhibit large deformations in response to applied stimuli. Local control of the orientation of LCEs spatially directs the deformation of these materials to realize spontaneous shape change in response to stimuli. Prior approaches to shape programming in LCEs utilize patterning techniques that involve the detailed inscription of spatially varying nematic fields to produce sheets. These patterned sheets deform into elaborate geometries with complex Gaussian curvatures. Here, we present an alternative approach to realize shape-morphing in LCEs where spatial patterning of the crosslink density locally regulates the material deformation magnitude on either side of a prescribed interface curve. We also present a simple mathematical model describing the behavior of these materials. Further experiments coupled with the mathematical model demonstrate the control of the sign of Gaussian curvature, which is used in combination with heat transfer effects to design LCEs that self-clean as a result of temperature-dependent actuation properties.

**Introduction**

Liquid crystalline elastomers (LCEs) are functional materials capable of generating large deformations in response to a stimulus.[1] The LCE deformation is associated with stimuli-induced disruption of order of the packing of liquid crystalline mesogens within the polymer network. For LCEs prepared with monodomain orientation, deformation takes the form of homogeneous contraction along the aligned axis of the material and expansion along the orthogonal axes.[2] In the last decade, material chemistries amenable to processing techniques have facilitated spatial programming of the local deformation of LCEs to realize complex deformations. Two approaches have been widely used for programming. The first method of producing deformation is through changing packing orientation of the liquid crystalline mesogens. Surface anchoring is commonly implemented as a technique to direct complex packing orientation. In this method, glass substrates are treated with a coating containing azobenzene units that are locally oriented with light.[3] Once monomers are filled into a cell fabricated using these coated substrates, the monomers take on the orientation of the coating and are subsequently polymerized. As a result, shapes such as cones or surfaces with Gaussian curvature are generated as order is disrupted between the mesogens.[4–13]

A second route to generating complex deformations is through localized control of actuation properties. While rather understudied in LCEs, this technique has been previously shown in numerous other materials to enable programming of complex shapes.[14–16] In this work we focus on encoding Gaussian curvature in the actuated state by designing interfaces between regions of prescribed properties that drive actuation. Such shape-morphing strategies use, for example, parallel strips of material with high and low equilibrium swelling factors. The dependence on film thickness, strip width, and swelling contrast have been studied in the context of thermally responsive photo-crosslinkable hydrogels and used to generate swelling-driven rolling patterns.[17,18] More recently, similar strategies have been extended to LCEs through the spatial patterning of gold nanoparticle concentration. The planar films experience out-of-plane buckling due to gradients in photothermal heat generation in response to uniform illumination.[19,20] However, this approach is dependent on the inclusion of nanoparticles within the LCE to localize a particular stimulus, such as temperature. Complex deformations can also be achieved via patterned illumination of a uniform LCE, but this requires precise control of the stimulus. [19,20] We propose to instead program localized properties into the LCE itself. Thus, in this work, we propose a method for generating complex shape transformations in monodomain LCEs consisting of two regions with distinct deformation responses based on inherent material properties.

Much like in other non-liquid crystalline systems, programming of crosslink density is an appealing route to introducing discontinuous actuation metrics. It has been established that properties relating to liquid crystalline behavior in these elastomers, particularly the nematic-to-isotropic transition temperature ($T_{NI}$) and the magnitude of deformation, are dependent upon the polymer network structure.[21–23] Given that LCEs are commonly fabricated by photopolymerization reactions, one way to locally control crosslink density is through spatial variation of light exposure time or intensity. However, this approach leaves regions of unreacted crosslinking monomers, which could be unstable and give rise to inconsistent material properties over time. Here, we report a technique for altering crosslink density by post-functionalization of LCEs. This approach is enabled by a thiol-acrylate photopolymerization that leaves pendant thiols attached to the backbone of the network.[24] We post-functionalize monodomain LCEs by reacting additional liquid crystalline diacrylate monomer with the pendant thiols in a second photopolymerization step. Using diacrylates for post-functionalization leads to an increase in crosslink density, and therefore, an increase in $T_{NI}$. We use the spatiotemporal control inherent in photoinitiation to locally pattern discrete regions of actuation metrics across various interface configurations and characterize the deformation in response to heat. While this technique could be used to program numerous spatial regions within a single film, here we focus only on patterning a single boundary between two regions of distinctive crosslink density. We find that Gaussian curvature develops across the interface between the two regions.

Using the experimental observations of actuation in spatially patterned LCEs with linear and curved interfaces between regions of distinct material properties, we developed a model to describe the resulting actuation. The result is a simple mathematical description that provides insight about the deformations seen in experiments. Simulated actuations were performed to confirm qualitative agreement between theory and experiment. Using insight from experiments coupled with theory, we discuss the programming of the sign of Gaussian curvature relative to experimental parameters, enabling a demonstration in which a dynamic LCE actuation is designed based on temperature-dependent phenomenon across the patterned materials.

**Experimental**

**Materials** 1,4-Bis-[4-(6-acryloyloxyhexyloxy)benzoyloxy]-2-methylbenzene (C6M) was purchased from Wilshire Technologies. 1,4-Benzenedimethanethiol (BDMT) was purchased from Tokyo Chemical Industry. Omnirad 819 [Bis(2,4,6-trimethylbenzoyl)phenylphosphine oxide)] was purchased from IGM Resins. 4-Methoxyphenol (MEHQ) was purchased from Sigma. Dichloromethane (DCM) was purchased from Fisher Scientific. All materials were used as received.

**Fabrication of Monodomain Liquid Crystalline Elastomers** The initial material was fabricated with C6M and BDMT in a 0.75:1 thiol to acrylate ratio, 1 wt% percent Omnirad 819, and 0.5 wt% MEHQ. The monomer mixture was melted at 150°C and vortexed. The melted mixture was filled into 30 μm alignment cells, fabricated by the antiparallel alignment of rubbed Elvamide-coated glass slides with 30 μm glass spacers, at 90°C (isotropic state). The filled alignment cells were cooled to 45°C (nematic state) and photopolymerized under UV light (365 nm, 50 mW/cm2) for 5 minutes.

**Post-Functionalization of Uniform and Patterned Films** The initial material was swelled in a DCM solution containing 33 mg/mL of C6M monomer and 0.67 mg/mL of Omnirad 819 for 24 hours. After being removed from the solution, the film dried for 24 hours. For uniform post-functionalized materials, the entire film was laid flat and polymerized under UV light (365 nm, 50 mW/cm2) at room temperature for 5 minutes. For patterned films, photomasks were printed with desired pattern and placed on top of the initial material to block incident light in select regions upon UV exposure. After post-functionalization, the LCEs were placed in pure DCM for 24 h to remove any remaining monomers and subsequently dried before further testing.

**Real-Time FTIR** Real-time Fourier transform infrared spectroscopy (FTIR) was performed using a Nicolet iS20 and a custom heating accessory. A sample of melted monomers was polymerized between NaCl salt plates at 45 °C and polymerized with 365 nm light (50 mW/cm2) for 5 min. Thiol (2500-2620 cm-1) and acrylate (790-820 cm-1) peak areas tracked during polymerization to calculate conversions.

**Tensile Testing** Tensile experiments were conducted by dynamic mechanical analysis (RSA-G2, TA Instruments). Materials were tested both parallel and perpendicular to the nematic director with strain applied at 5%/min. The moduli of the LCEs were taken in the linear regime of the stress-strain curve between 2 and 4% strain.

**Thermomechanical Actuation** LCE strips (~2 mm x 10 mm x 0.03 mm) cut along the nematic director were heated at 5 ºC/min from 25-225ºC while held in tension (constant force of 0.005 N).

Nematic to isotropic transition temperature was calculated as the inflection point of the strain curve.

**Differential Scanning Calorimetry** Differential scanning calorimetry (Discovery DSC 2500, TA Instruments) was used to measure glass transition temperatures of the LCEs. Measurements were recorded from -50 to 150 ºC at 5 ºC/min. Second heating cycles were used for analysis.

**Gel Fraction and Monomer Uptake** Gel fractions were measured for the initial material by placing the material in dichloromethane for 24 h followed by 24 h of drying time. Gel fractions were calculated from the ratio of the initial mass and the mass of the dried polymer.

Monomer uptake was measured by swelling a sample of initial material in a dichloromethane solution containing 33 mg/mL of C6M monomer and 0.67 mg/mL of Omnirad 819 for 24 h followed by 24 h of drying time. The ratio of the mass of the dried polymer to the mass of the initial polymer was calculated and compared to the gel fractions to determine monomer uptake as a percentage of the initial film mass.

**Dynamic LCE Demonstration** An LCE was patterned with five circular photomask regions as described in the previous sections. The material was secured with Kapton tape to a glass slide coated with graphite. Small rocks were placed on the sample and the entire configuration was placed on the surface of a hot plate heated to 180°C and tilted to an angle of 30°.

**Results & Discussion**

**Post-Functionalization of LCEs** LCEs were prepared by the polymerization of the liquid crystalline monomer (C6M) and benzenedimethanethiol (BDMT) in a 0.75:1 thiol:acrylate ratio (Figure 1a). These monomers were mixed with photoinitiator (Omnirad 819, 1 wt%) in the isotropic state, filled into 30 μm alignment cells, cooled to the nematic state, and polymerized using 365 nm light (5 min, 50 mW/cm$^2$). As described in a recent examination,[24] LCEs prepared by free-radical chain transfer reactions retain high concentrations of unreacted thiols. These thiols were post-functionalized by swelling in additional C6M/photoinitiator and subjecting to photopolymerization. The post-functionalization of the LCE increased the crosslink density (Figure 1b) and resulted in an approximately 10% increase in mass. Thiol and acrylate conversion during uniform bulk polymerizations are shown for each step in Figure S1. Excess monomer was removed following post-functionalization by swelling in pure DCM and drying. As shown in WAXS diffraction patterns presented in Figure S2 and order parameters in Table 1, the orientation direction and magnitude of the nematic director were retained after post-functionalization. Material properties were compared for homogeneous films as described in Table 1. The post-functionalized LCE had a higher glass transition temperature and elastic moduli (both parallel and perpendicular to the nematic director) than the initial as-prepared LCE. Most notably, as evident in Figure 1c, post-functionalization of the LCE shifted the thermomechanical response to higher temperatures than the as-prepared LCE, corresponding to nematic-to-isotropic transition temperatures ($T_{NI}$) of 203°C and 181°C, respectively. Therefore, at any given temperature within this actuation range, there is a discrepancy between the strain generation for the two differently crosslinked materials. Further discussion will refer to actuation strains in the as-prepared material as $\lambda_1$ and the post-functionalized, more crosslinked material as $\lambda_2$, where $\lambda$ is the length of the LCE at a given temperature as a fraction of the original length along the nematic director.

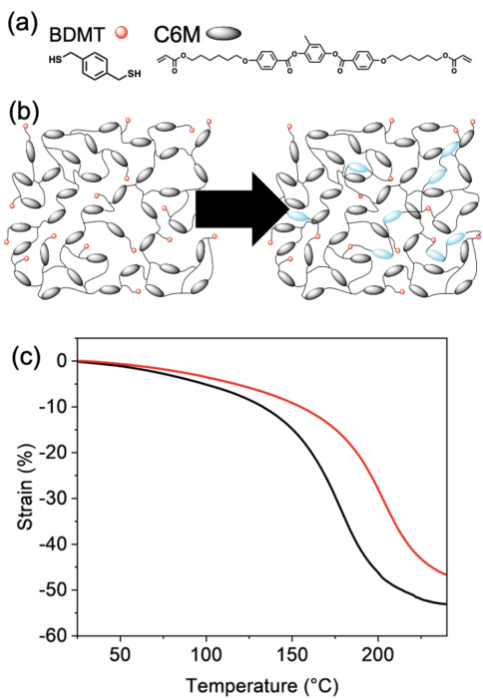

**Figure 1.** Crosslink density variation changes thermomechanical response. (a) LCEs were prepared by copolymerization of benzenedimethanethiol (BDMT) and C6M. (b) The free-radical mediated chain transfer reaction results in an LCE typified by a polymer network composed of liquid crystalline mesogens (gray ellipses) and pendant thiols (red circles). The unreacted thiols were post-functionalized by swelling in additional acrylate monomer (blue ellipses) that increased the number of crosslinks upon photopolymerization. (c) The thermomechanical response of the LCE as prepared (black) and after post-functionalization (red) is distinguished over a 25-250°C temperature ramp (at 5 °C/min).

Table 1. Material properties for the initial and post-functionalized LCEs.

| | $S^{(a)}$ | $T_g^{(b)}$ (°C) | ∥ Modulus$^{(c)}$ (MPa) | ⊥ Modulus$^{(c)}$ (MPa) | $T_{ni}^{(d)}$ (°C) | λ at 180°C$^{(e)}$ |
|---|---|---|---|---|---|---|
| As-Prepared | 0.57 | 8.5 | 35 ± 6 | 7.3 ± 0.2 | 181 ± 3 | 0.72 ± 0.03 |
| Post-Functionalized | 0.61 | 13.5 | 56 ± 5 | 9.2 ± 0.2 | 203 ± 9 | 0.82 ± 0.01 |

$^{(a)}$ Herman's orientation parameter, calculated from WAXS diffraction data (Figure S2)

$^{(b)}$ Reported value from midpoint in second DSC heating cycle (5°C/min, Figure S3)

$^{(c)}$ Tensile pull 5%/min, modulus calculated in linear strain regime (2-4%, Figure S4)

$^{(d)}$ Calculated as maximum derivative of thermomechanical strain generation (temp ramp 5°C/min) in the range of 25-225°C

$^{(e)}$ Calculated as ratio of length along nematic director at 180°C to length at 25°C

**Patterning of Actuation Properties** LCEs were then prepared with discontinuous actuation metrics using spatial patterning of the crosslink density by local post-functionalization. The local post-functionalization of the LCE was achieved by swelling the as-prepared LCE with the solution of DCM, C6M, and I819, drying of the film, and subsequent UV irradiation through a photomask printed with the desired pattern. Unreacted monomer was washed from the film with DCM after post-functionalization and the materials were allowed to dry. As an initial demonstration, an LCE was prepared with a straight interface between regions of $\lambda_1$ and $\lambda_2$ (Figure 2a). This interface between the two patterned regions intersected the nematic director at an angle of 90°. Upon exposure to elevated temperature at which there was large separation between the documented $\lambda_1$ and $\lambda_2$ (180 °C), Gaussian curvature was evident in the deformation (Figure 2b, 2c).

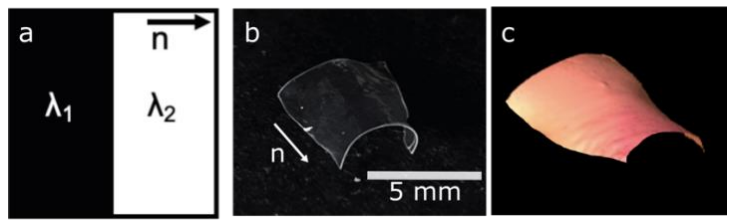

**Figure 2.** (a) Schematic illustrating masked (black) and unmasked (white) regions of monodomain LCEs prepared by local post-functionalization. Photograph (b) and topographical image (c) of the patterned LCEs are presented when the material was heated to 180°C

Next, LCEs were patterned with curved interfaces between regions of $\lambda_1$ and $\lambda_2$. Illustrated in Figure 3a, a circular arc of radius 0.75 cm was designated as the interface curve. The linear nematic director was patterned horizontally in the film. Upon actuation at 180°C, the LCE deformed to produce the shape illustrated in Figure 3b and 3c. This deformation appeared similar in shape to the actuation with the linear interface in Figure 2. Probing further with curved interfaces, we patterned an LCE with a full circular boundary between the two regions of contrasting crosslink density, resulting in a $\lambda_1$ region contained inside a circle of radius 0.25 cm and a $\lambda_2$ region surrounding the circle as shown in Figure 3d. Again, a linear nematic director was patterned horizontally in the square LCE. Upon heating to 180°C, the circular pattern resulted in a considerably more complex deformation, shown in Figures 3e and 3f.

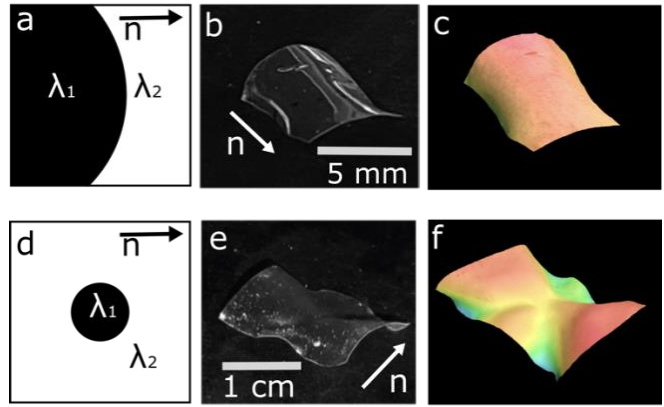

**Figure 3**. Schematics illustrate masked (black) and unmasked (white) regions of monodomain LCEs during post-functionalization exposure for (a) a transverse arc interface and (d) a centered circular interface. Photographs (b,e) and topographical images (c,f) of the patterned LCEs are presented when the material was heated to 180°C

**Theory and Model Development** To capture qualitative features of the observed experiments and provide a basis for generating further designs, we developed a simplified mathematical model. The alignment direction in the LCE was modeled by a normalized director field $n$. We assumed a 2D thin sheet model with prescribed thickness *t*, adopted a Cartesian coordinate system *(x, y)*, and assumed that the director was uniform and aligned with the *x*-axis. It was assumed that the domain $D$ was split into two regions separated by a smooth curve, such that $\lambda = \lambda_1$ in one region and $\lambda = $ $\lambda_2$ in the other, where $\lambda_1$ and $\lambda_2$ were viewed as constants for a given stimulus parameter (e.g., temperature). The associated actuation metric of the system then adopts the form

$$ds^2 = \lambda(x,y)^2 dx^2 + \lambda(x,y)^{-2\nu} dy^2 \qquad [1]$$

where $\lambda = \lambda(x,y)$ is a smooth *effective* contraction factor that emerges from a monotonic transition from $\lambda_1$ to $\lambda_2$ and $\nu$ is the opto-thermal Poisson ratio that is equal to $1/2$ in the case of elastomers but can be as high as $\nu = 2$ in glassy liquid crystalline polymer networks. The model's assumptions were derived from experimental observations that show the formation of smooth surfaces with zero Gaussian curvature away from the interfacial curves. These observations suggested that any developed intrinsic curvature was the result of a rapid variation in the deformation factor $\lambda$ concentrated in a narrow width around the boundary between the two regions. As such, we defined $2\Delta$ as a transition width parameter centered on the unactuated boundary. To ensure that this model yielded a well-defined metric, we assumed that the $\Delta$ was smaller than the minimum radius of curvature of the interface and used a smooth interpolating function $S(\cdot, \lambda_1, \lambda_2)$ that monotonically transitions from $\lambda_1$ to $\lambda_2$ over a finite interval $(-1,1)$. The effective contraction factor could then be written as

$$\lambda(x,y) = S(s(x,y)/\Delta, \lambda_1, \lambda_2) \qquad [2]$$

where $s(x,y)$ is the signed distance from the point $(x,y)$ to the interfacial curve such that $s(x,y) < 0$ if $(x,y)$ lies in the region of $\lambda_1$. For further details, see the Supplementary Information.

To test the applicability of this model, we used it to compute the 2D Gaussian curvature profiles for the arc interface and circular interface examples that were tested experimentally. Furthermore, by modeling each system as a thin sheet of incompressible Neo-Hookean elastomer, we computed their activated 3D equilibrium configurations by numerically minimizing standard stretch and bend elastic shell energies using MorphoShell.[25,26] The simulations were run with material parameters and actuation conditions that matched those used in experiments. Both the 2D Gaussian curvature profiles and 3D simulation results are shown in Figure 4.

We note that the same Gaussian curvature pattern was observed in the 2D and 3D computations, with the notable difference that in the 3D computation the Gaussian curvature was diminished and spread over a wider neighborhood of the interfacial curve due to the non-isometric effects of the bend energy, which was expected based on prior demonstrations. [25,26] While the simulations verified qualitative agreement between experiment and theory, we note that precise quantitative agreement of extrinsic geometry across the entirety of the sheet was not achieved given the sensitivity of the experiments to factors such as thermal and mechanical interactions with the resting surface. However, features such as characteristic sign changes of Gaussian curvature along and across the interfacial curves and vanishing of Gaussian curvature away from the boundary were in good agreement. Therefore, the model described here does explain the key differences seen in the LCEs as a function of the interfaces between patterned regions of actuation properties. Both our model and experiments showed that rich curved topographies can emerge from such relatively simple systems even though the Gaussian curvature is concentrated in a narrow neighborhood of the patterned interfacial curves.

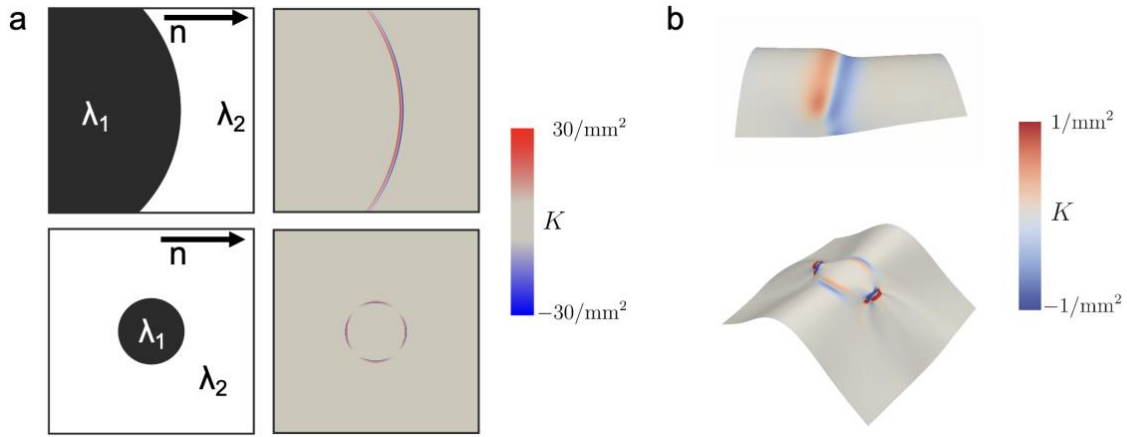

**Figure 4.** (a) Schematics illustrate masked (black) and unmasked (white) regions of monodomain LCEs during post-functionalization exposure and corresponding Gaussian curvature profiles computed directly from the 2D metric for $\Delta = 0.09$ mm. (b) Simulations for corresponding 3D equilibrium configurations and Gaussian curvature $K$ computed for sheets of thickness $t = 0.03$ mm (right).

**Curvature generation along linear interfaces** Informed by both experiment and theory, we expected the magnitude and sign of the Gaussian curvature near the interface curve to depend on the angle at which the nematic director intersects the patterned interface. Using a heuristic argument, we predicted that if the director and the interface curve were to intersect at a particular angle where the interface curve is deformed by equal amounts on both sides (θ*), then the Gaussian curvature generated at the point of intersection would be minimized and approximately zero. To derive the angle θ* based on the mathematical model described in the previous section, we considered a uniform director field that makes an angle θ with a linear interface between the two regions. As the LCE is actuated, the interface's target squared length element is given by $ds_i^2 = \cos^2\theta \lambda_i^2 + \sin^2\theta/\lambda_i$ for $i = 1, 2$, on either side of the interface, where we have taken $\nu = 1/2$. To avoid the generation of Gaussian curvature, we require $ds_1^2 = ds_2^2$ which yields

$$\tan^2\theta = \lambda_1 \lambda_2 (\lambda_1 + \lambda_2) \qquad [3]$$

Equation 3 is satisfied by a unique $\theta^* \in (0, \pi/2)$, which for the experimental system's values is approximately 43°. Accordingly, rectangular LCEs were prepared with diagonal linear interfaces between patterned regions of crosslink density as shown in Figure 5a with values of θ spanning 41-46° (Figure 5b). Upon exposure to heat (180°C), the pattern with a θ angle of 43° did indeed result in a deformation exhibiting zero Gaussian curvature, as predicted by the formula in Equation 3. Furthermore, we also performed simulations to compute 3D equilibrium configurations of activated square sheets with angles spanning a broader range of θ. The simulations were performed at various angles ranging from 0° to 90° (Figure 5d), with a particular focus around $\theta^* = 43°$ (Figure 5c) to validate the predictions based on Equation 3.

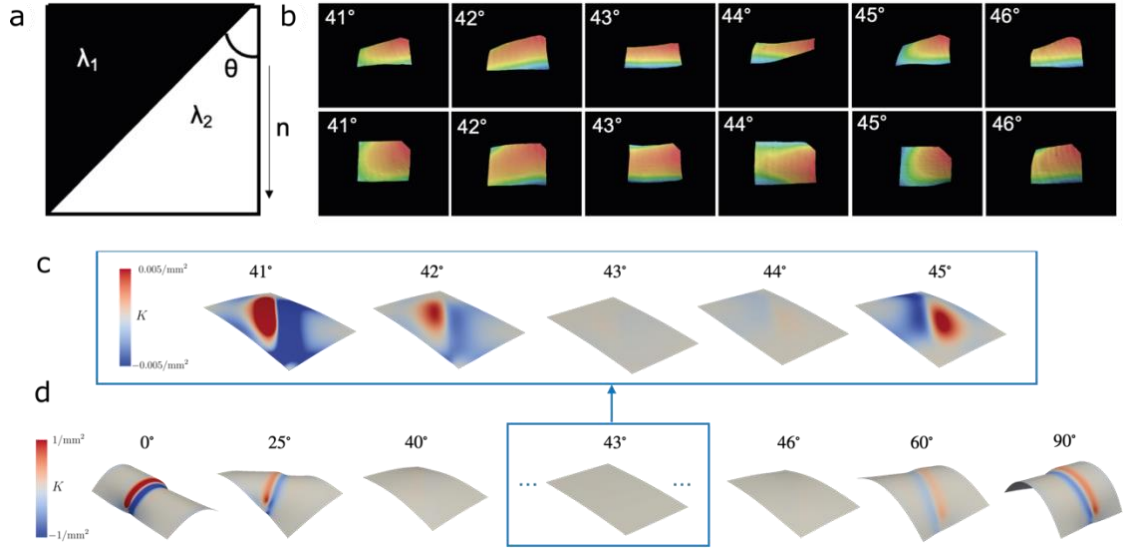

**Figure 5.** (a) LCEs were prepared with straight diagonal interfaces with specified angle, θ, relative to the nematic director. (b) The deformations of the patterned LCE samples are shown from the side (top row) and top-down (bottom row) for values of θ (indicated in each image) bridging the region where Gaussian curvature is inverted around 43°. Computational simulations of equilibrium configurations and Gaussian curvature $K$ with parameters $\Delta = 0.09$ mm and thickness $t = 0.03$ mm for activated square sheets with straight interfaces intersecting a uniform director field at various angles (c) emphasizing a particular focus on angles near 43° as a subset of (d) angles from 0° to 90°.

**Programming Dynamic Actuation** While the initial zero-curvature analysis was done for a particular set of experimental conditions, it should be noted that the calculation of θ in Equation 3 is dependent on values of $\lambda$. Since $\lambda$ is temperature-dependent, the zero-curvature angle $\theta^*$ is also temperature dependent. Using this temperature dependency as an advantage, active surfaces were developed based on patterning of discontinuous metrics. Specifically, an LCE with a uniform nematic director was patterned with photomasks in 5 circular regions. As illustrated in Figure 6a, four points exist around each circle where the angle θ between the nematic director and patterned interface should result in zero Gaussian curvature. Therefore, as this material is heated, five regions of complex deformation emerge. However, when left open to the air (22°C) and heated from only the bottom surface (180°C), the material experienced thermal fluctuations as it deformed away from the surface. These fluctuations resulted in shifting of the value of θ for zero Gaussian curvature, and accordingly, the deformation oscillated as the sign of the Gaussian curvature fluctuated between positive and negative across these regions (Figure 6b).

The spontaneous interconversion of curvature in these dynamic LCEs could be utilized as a "self-cleaning" surface, useful in applications ranging from consumer goods to solar panels. Notably, the LCEs examined here are low haze and do not absorb wavelengths of light that are typically harnessed by solar panels (Figure S5). As an illustration of self-cleaning, an LCE was prepared with the pattern shown in Figure 6a. Small rock particles were placed on the patterned material at ambient temperature. The system was then tilted at an angle of 30°, as many solar panels are designed, and heated from the bottom, emulating the surface of a solar panel becoming hot as it absorbs sunlight. As the patterned LCE was heated, the material deformation resulted in the formation of thermal gradients due to loss of contact with the heat source. Accordingly, the material dynamically fluctuated through active values of θ and dislodged the contaminants (Figure 6c, Video S1).

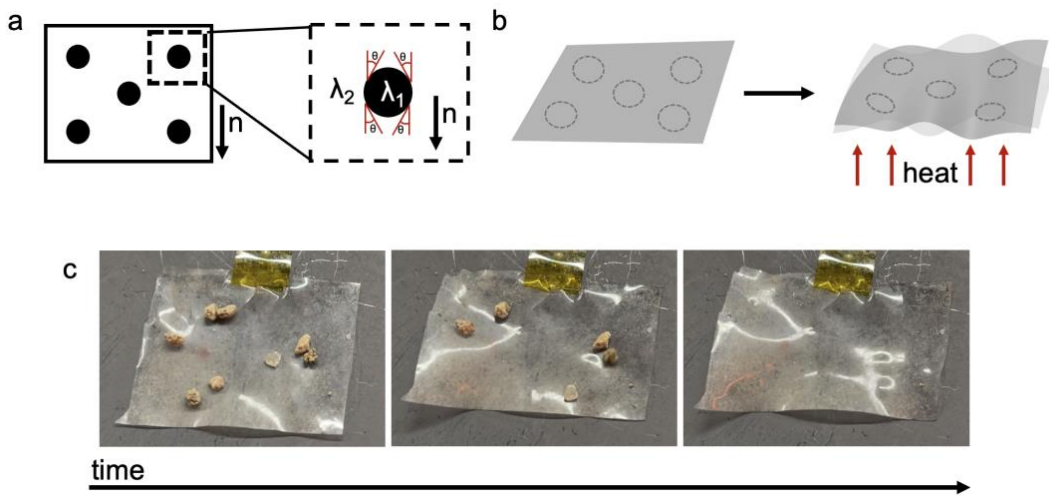

**Figure 6.** (a) LCEs were patterned with five circles, such that four angles exist around each circle where the Gaussian curvature profile flips. (b) Upon heating, the material deforms, and the shape dynamically changes as the material loses contact with the heating surface. The resulting deformation fluctuates as the actuation parameters vary with the local temperature. (c) The dynamic shape changes of the LCE removed debris from its surface.

**Conclusions**

Post-functionalization of LCEs was implemented to spatially pattern crosslink density in LCEs using photopolymerization. Patterned LCEs prepared with spatial variation in crosslink density exhibited regions of distinct actuation metrics separated by a prescribed interface. Upon heating, the LCEs exhibited Gaussian curvature in their deformation. Computational analysis confirmed the evolution of curvature as related to actuation discrepancies across the patterned interface. Overall, this study presents a distinctive approach to preparing LCE with a uniform director field capable of spatial variation in crosslink density to realize complex shape change. One potential functional use is illustrated, as a self-cleaning surface, but others are envisioned, such as in optical materials in which it is highly desirable and necessary to decouple shape change from optical birefringence, especially in patterned lenses.

ASSOCIATED CONTENT

**Supporting Information**.

The following files are available free of charge.

PDF containing additional computational details, FTIR data, WAXS data, DSC data, tensile data, and UV-Vis data

AUTHOR INFORMATION

**Corresponding Author**

Timothy J. White − Department of Chemical and Biological Engineering and Materials Science and Engineering Program

University of Colorado Boulder, Boulder, Colorado 80309, United States;

Email: timothy.j.white@colorado.edu

**Author Contributions**

T.S.H, C.M., I.G, C.N.B., and T.J.W. proposed and designed the research project. T.S.H. and R.G.A.B. performed the experiments. C.M., I.G., and D.D. undertook theoretical calculations and analyzed data. D.D. performed the simulations and computational modeling. All authors contributed to the development of ideas and preparation of the paper.

ACKNOWLEDGMENT

The authors acknowledge helpful conversations with Ryan C. Hayward. T.S.H. acknowledges Graduate Research Fellowship support from the National Science Foundation. D.D. is supported by the EPSRC Centre for Doctoral Training in Computational Methods for Materials Science, grant No. EP/L015552/1. C.M. was supported by a Presidential Postdoctoral Fellowship at NTU and an Early Career Research Fellowship at Fitzwilliam College, Cambridge.

# Supplementary Information

# Discontinuous metric programming in liquid crystal elastomers

Tayler S. Hebner,[a] Riley G.A. Bowman,[a] Daniel Duffy,[b] Cyrus Mostajeran,[b,c] Itay Griniasty,[d] Itai Cohen,[d] Mark Warner,[e] Christopher N. Bowman,[a,f] and Timothy J. White*[a,f]

[a] Department of Chemical and Biological Engineering, University of Colorado Boulder, 596 UCB, Boulder, CO 80309, USA
[b] Department of Engineering, University of Cambridge, Cambridge, England CB2 1PZ, United Kingdom
[c] School of Physical and Mathematical Sciences, Nanyang Technological University, Singapore 637371
[d] Laboratory of Atomic and Solid State Physics, Cornell University, Ithaca, New York 14853-2501, USA
[e] Cavendish Laboratory, University of Cambridge, Cambridge CB3 0HE, United Kingdom
[f] Materials Science and Engineering Program, University of Colorado Boulder, 596 UCB, Boulder, CO 80309, USA

**Model Development** The alignment direction in the LCE was modeled by a normalized director field n. We assumed a 2D thin sheet model with prescribed thickness *t*, adopted a Cartesian coordinate system *(x, y)*, and assumed that the director is uniform and aligned with the $x$-axis. The associated actuated metric takes the form $ds^2 = \lambda^2 dx^2 + \lambda^{-2\nu} dy^2$ where $\lambda < 1$ is a contraction factor and ν is the opto-thermal Poisson ratio that is equal to $1/2$ in the case of elastomers but can be as much as $\nu = 2$ in glassy liquid crystalline polymer networks. It was assumed that the domain $D$ was split into two regions, $\Lambda_1$ and $\Lambda_2$, such that $\lambda = \lambda_1$ in $\Lambda_1$ and $\lambda = \lambda_2$ in $\Lambda_2$, where $\lambda_1$ and $\lambda_2$ were viewed as constants for a given stimulus parameter (e.g., temperature).

We began with the simplest example where the regions are separated by the $y$-axis so that $\Lambda_1 = \{(x,y) \in D: x < 0\}$ and $\Lambda_2 = \{(x,y) \in D: x > 0\}$. Assuming that these two regions are connected along the interfacing curve, the actuated metric of the resulting system is modeled by

$$ds^2 = \lambda(x,y)^2 dx^2 + \lambda(x,y)^{-2\nu} dy^2 \qquad [1]$$

where

$$\lambda(x,y) = \lambda_1 + \frac{\lambda_2 - \lambda_1}{2}\left(\tanh\frac{x}{\Delta} + 1\right) \qquad [2]$$

and $\Delta$ is a transition width parameter dependent on the fabrication process, which controls the rate of the smooth and monotonic transition from $\lambda_1$ to $\lambda_2$ described by Equation (2).

The model was then generalized to systems with nonlinear interface curves between $\Lambda_1$ and $\Lambda_2$ by taking a normal tubular neighborhood of width $2\Delta$ of the interface curve and smoothly transitioning from $\lambda_1$ to $\lambda_2$ across it. Given a planar interface curve $C$ separating regions $\Lambda_1$ and $\Lambda_2$, the $\Delta$-neighborhood of $C$ is generated by non-intersecting straight-line segments normal to the curve extending by a length of $\Delta$ either side of the curve. A necessary condition for such a region to be well-defined is that

$$\Delta < \frac{1}{sup_{p \in C} k(p)} \qquad [3]$$

where $k(p)$ is the curvature of $C$ at $p$. That is, we require that $\Delta$ is smaller than the minimum radius of curvature of $C$, which is a practically reasonable assumption. Locally, this condition is also sufficient.

Note that even if $\Delta$ is sufficiently small to yield a well-defined tubular transition region around $C$, we cannot use a typical sigmoid function such as (2) to obtain a smoothly varying metric. The resulting $\lambda(x, y)$ will generally be ill-defined at points of intersection of normals to $C$ beyond the $\Delta$-neighborhood since the value of $\lambda$ will generally depend on the choice of normal used to extend it. To get around this, we use monotonic interpolating functions $S$ that transition from $\lambda_1$ to $\lambda_2$ over a finite interval. Specifically, we seek monotonic functions $S = S(z)$ that satisfy $S(z) = \lambda_1$ for $z \leq -1$, $S(z) = \lambda_2$ for $z \geq 1$, $S'(-1) = S'(1) = 0$, and $S''(-1) = S''(1) = 0$. The piecewise quintic polynomial $S(z; \lambda_1, \lambda_2)$ given by $S(z; \lambda_1, \lambda_2) = \lambda_1$ for $z \leq -1$, $S(z; \lambda_1, \lambda_2) = \lambda_2$ for $z \geq 1$, and

$$\frac{3}{16}(\lambda_2 - \lambda_1)z^5 - \frac{5}{8}(\lambda_2 - \lambda_1)z^3 + \frac{15}{16}(\lambda_2 - \lambda_1)z + \frac{1}{2}(\lambda_1 + \lambda_2)$$

for $-1 < z < 1$ is monotonic and satisfies the required properties. We used this function in the simulations presented in the next section.

In the case where the normal $\Delta$-neighborhood $\mathcal{N}_\Delta \subset D = \Lambda_1 \cup C \cup \Lambda_2$ of the interface curve $C$ is well-defined, we can model the target metric of the system as $ds^2 = \lambda(x, y)^2 dx^2 + \lambda(x, y)^{-2\nu} dy^2$, where

$$\lambda(x, y) = \begin{cases} \lambda_1 & if \quad (x, y) \in \Lambda_1 \setminus \mathcal{N}_\Delta \\ S(-|s|/\Delta; \lambda_1, \lambda_2) & if \quad (x, y) \in \Lambda_1 \cap \mathcal{N}_\Delta \\ S(|s|/\Delta; \lambda_1, \lambda_2) & if \quad (x, y) \in \Lambda_2 \cap \mathcal{N}_\Delta \\ \lambda_2 & if \quad (x, y) \in \Lambda_2 \setminus \mathcal{N}_\Delta \end{cases} \qquad [4]$$

and $s = s(x, y)$ denotes the smallest distance of point $(x, y)$ to the curve $C$ in the unactuated state.

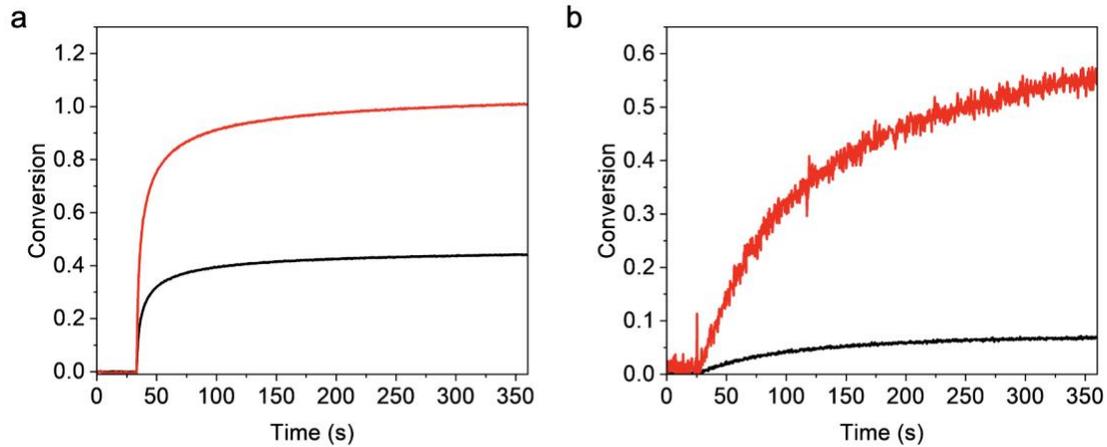

**Fig. S1**

Real-time FTIR data for thiol (black) and acrylate (red) conversion during photopolymerization in the **a** initial fabrication and **b** post-functionalization steps. Samples were exposed to UV light after 30 seconds (365 nm, 50 mW/cm$^2$)

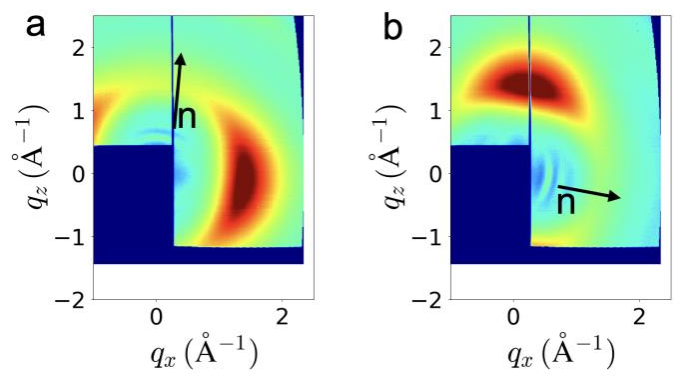

**Fig. S2** WAXS diffraction patterns for **a** initial and **b** post-functionalized materials.

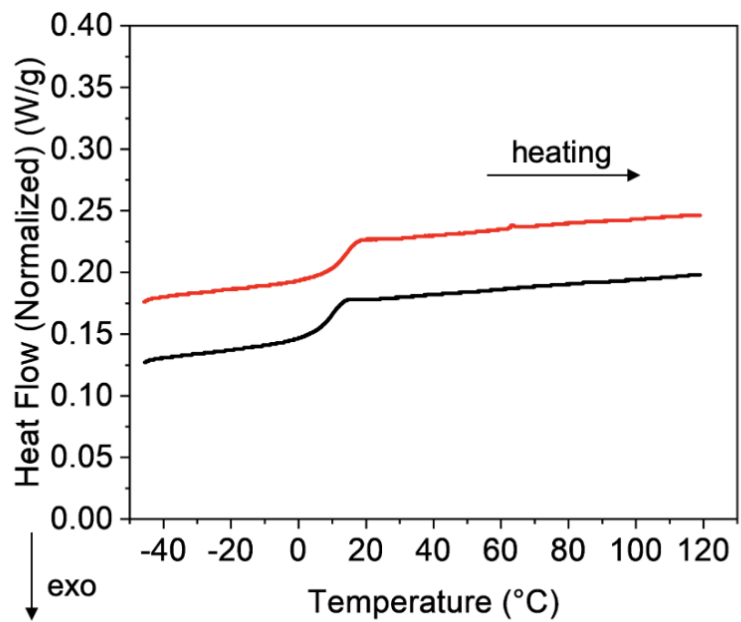

**Fig. S3** DSC thermograms for initial (black) and post-functionalized (red) materials as temperature was ramped from -50°C to 120°C at 5°C/min. Second heating curves used for analysis.

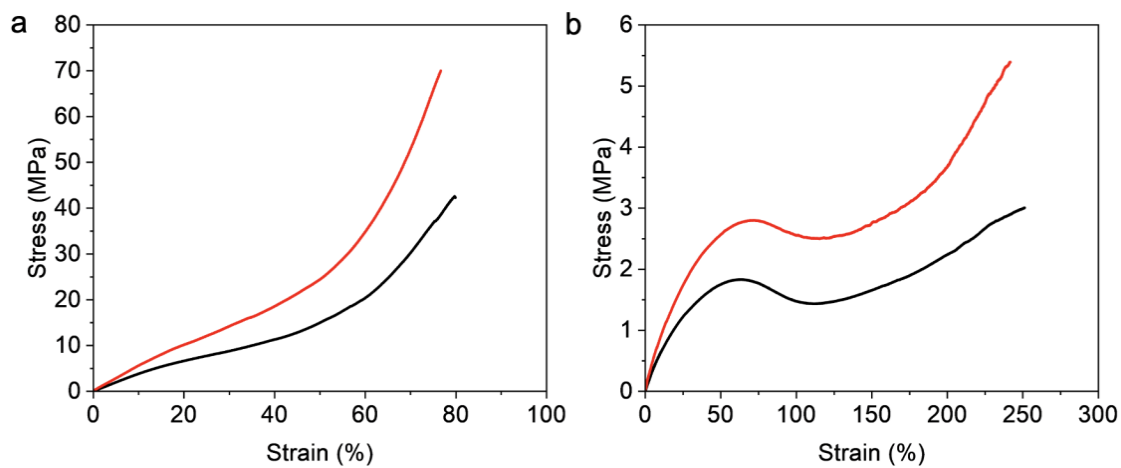

**Fig. S4** Tensile data (5%/min) for initial (black) and post-functionalized (red) materials pulled **a** parallel to the nematic director and **b** perpendicular to the nematic director

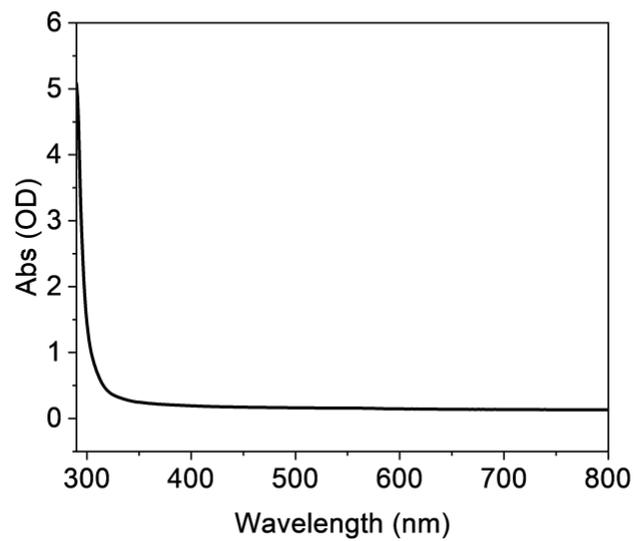

**Fig. S5** UV-Vis spectrum for the LCEs used in this study